\documentclass[preprintnumbers,amsmath,amssymb,floatfix,12pt,prd,superscriptaddress,nofootinbib]{revtex4}
\usepackage{graphicx}
\usepackage{epsfig}
\usepackage{bm}
\usepackage{amsfonts}
\usepackage{subfigure}

\def\bea{\begin{eqnarray}}
\def\eea{\end{eqnarray}}

\begin{document}
\begin{center}
\LARGE {\bf Cosmological perturbations in warm-tachyon inflationary universe model with viscous pressure on the brane}
\end{center}
\begin{center}
{M. R. Setare $^{a}$\footnote{E-mail: rezakord@ipm.ir
}\hspace{1mm} ,
V. Kamali $^{a}$\footnote{E-mail: vkamali1362@gmail.com}\hspace{1.5mm} \\
 $^a$ {\small {\em  Department of Science, University of Kurdistan \\
Sanandaj, IRAN.}}}\\

\end{center}


\begin{center}
{\bf{Abstract}}\\
We study warm-viscous inflationary universe model on the brane, in a tachyon field theory. We obtain the general conditions
which are required for this model to be realizable. In longitudinal gauge, the primoradial perturbation parameters are found in great details,
using slow-roll and quasi-stable approximations. The general expressions of the tensor-to-scalar ratio, scalar spectral index and its running are found.
We derive the characteristics of the inflationary universe model by using an effective exponential potential in two cases: 1- Dissipative parameter $\Gamma$ and bulk viscous parameter $\zeta$ are constant parameters.
2- Dissipative parameter  as a function of tachyon field $\phi$ and bulk viscous parameter
as a function of radiation-matter mixture  energy density $\rho$. The parameters of the model are restricted by
recent observational data from the seven-year Wilkinson microwave
anisotropy probe (WMAP7).
 \end{center}

\newpage

\section{Introduction}
As is well-known, the Big Bang model have many long-standing problems (horizon,
flatness,...). These problems may be  solved in the framework of the
inflationary universe models \cite{1-i}.
The causal interpretation of the origin of
the distribution of Large-Scale Structure (LSS) and observed anisotropy
of cosmological microwave background (CMB) \cite{6} are provided by the scalar field as a source
of inflation. For standard
models of inflationary universe (cool inflation models), the inflation period may be divided
into two regimes, slow-roll and reheating epochs. Firstly, in slow-roll
period kinetic energy remains small compared to the potential energy
term. In this period, all interactions between scalar fields
(inflatons) and  other fields are neglected and the universe
inflates. Subsequently, in reheating epoch, the kinetic segment of energy
is comparable to  the potential energy and the inflaton field starts an
oscillation around the minimum of the potential by losing their
energy to other fields which are  present in the theory. Therefore, the reheating epoch
is the end period of inflation. In cool inflation models the quantum fluctuations may be responsible for Large-Scale Structures  formation. \\ Radiation production in warm inflation models
occurs during inflationary period and
reheating period may be avoided \cite{3}. Thermal fluctuations are
obtained during warm inflation. Thermal fluctuations instead of quantum fluctuations  could play a
dominant role to produce initial fluctuations which are necessary
for Large-Scale Structure (LSS) formation. Therefore the density
fluctuations arise  from thermal rather than quantum fluctuations
\cite{3-i}. Finally,  warm inflationary period ends when the universe stops
inflating. After the warm inflation period the universe enters in radiation
phase smoothly \cite{3}. Remaining inflatons or dominant
radiation fields could create  the matter components of the universe. Some extensions of this model are found in Ref.\cite{new}.\\
In warm inflation models, for simplicity, the particles which are created by the inflaton decay are considered as massless particles (or radiation). The existence of massive particles in the inflationary fluid model as a new model of inflation have  considered in Ref.\cite{41-i}.
The perturbation parameters of this model have obtained in Ref.\cite{2-ne}.
In this scenario the existence of massive particles may  altere the dynamics of the inflationary universe model by modification the fluid pressure.
Decay of the massive particles whitin the fluid is an entropy-producing scalar phenomenon and in other hand
"bulk viscous pressure" have entropy-producing property.
Therefore the decay of particles  may be considered by a bulk viscous pressure $\Pi=-3\zeta H$ \cite{3-ne} where $H$ is
Hubble parameter and $\zeta$ is phenomenological coefficient of bulk viscosity.
This coefficient is positive-definite by the second law of thermodynamics and in general depends on the energy density of the fluid.\\
In  higher dimensional theories, Friedmann-Robertson-Walker (FRW) cosmological models in the
context of string/M-theory have related to brane-antibrane
configurations \cite{4-i}. It was shown that the tachyon fields associated with
unstable D-branes may be responsible for inflation in early time
\cite{5-i}. Tachyonic inflationary  universe model is a k-inflation model
\cite{n-1}, for the  scalar field $\phi$ with a positive potential
$V(\phi)$. Tachyonic form of effective potential has two special properties,
firstly the  maximum of this  potential is found where
$\phi\rightarrow 0$ and second property is the minimum of the
potential is obtained where $\phi\rightarrow \infty$. In cool inflation models, if the
tachyonic  field starts to roll down the potential, then universe which is
dominated by a new form of matter will smoothly evolve from
inflationary universe to an era  dominated by a
non-relativistic fluid \cite{1}. We will explain the phase
of acceleration expansion (inflation) in term of tachyon field in the context of warm inflation theory.\\
We may live on a brane which is embedded in a higher
dimensional universe. This realization has significant implications to cosmology
\cite{1-f}. In this scenario, which is motivated by string theory, gravity (closed string modes)
can propagates in the bulk, while the standard model of particles (matter fields which are related to open string modes) is confined
to the lower-dimensional brane \cite{2-f}. $4d$ Einstein's equation projected onto the brane have been found in Ref.\cite{3-f}. Friedmann equation and the equations of linear perturbation theory \cite{4-f} may be modified by these projections. We would like to study the warm tachyon inflation model in the brane scenario. Therefore we will consider the linear cosmological perturbations theory for warm tachyon inflation model on the brane.  Einstein's equations which are projected onto the brane with cosmological constant and matter fields which are confined to 3d-brane, have the following form \cite{3-f}
\begin{equation}\label{1}
G_{\mu\nu}=-\Lambda_4 g_{\mu\nu}+(\frac{8\pi}{M_4^2})T_{\mu\nu}+(\frac{8\pi}{M_5^2})^2\pi_{\mu\nu}-E_{\mu\nu}
\end{equation}
where $E_{\mu\nu}$ is a projection of 5d Weyl tensor, $T_{\mu\nu}$ is energy density tensor on the brane, $\pi_{\mu\nu}$ is a tensor  quadratic in $T_{\mu\nu}$, $M_4$ and $M_5$ are Planck scales in 4 and 5 dimensions, respectively. Cosmological constant $\Lambda_4$ on the brane in term of 3-brane tension $\lambda$ and $5d$ cosmological constant $\Lambda$ is given by
\begin{equation}\label{}
\nonumber
\Lambda_4=\frac{4\pi}{M_5^3}(\Lambda+\frac{4\pi}{3M_5^3}\lambda^2)
\end{equation}
$4d$ Planck scale is determined by $5d$ Planck scale as
\begin{equation}\label{}
\nonumber
M_4=\sqrt{\frac{3}{4\pi}}(\frac{M_5^2}{\sqrt{\lambda}})M_5
\end{equation}
In spatially flat FRW model Friedmann equation, by using Einstein's equation (\ref{1}), has the form \cite{3-f}

\hspace{1.5mm}
\begin{equation}\label{i1}
H^2=\frac{\Lambda_4}{3}+(\frac{8\pi}{3M_4^2})\rho_T+(\frac{4\pi}{3M_5^3})\rho_T^2+\frac{\varepsilon}{a^4}
\end{equation}
where $a$ is scale factor of the model and $H$ is Hubble parameter $\varepsilon$ is an integration constant and $\rho_T$ is the total energy density on the brane. During inflation we could neglect $\Lambda_4$ and $\varepsilon$ constants. Therefore the Friedmann equation reduced to
\begin{equation}\label{}
\nonumber
H^2=\frac{8\pi}{3M_4^2}\rho_T(1+\frac{\rho_T}{2\lambda})
\end{equation}
In warm-viscous inflationary models where, the total energy density $\rho_T=\rho_{\phi}+\rho$ is found on the brane \cite{5-f}, the  Friedmann equation have the form
\begin{equation}\label{}
\nonumber
H^2=\frac{8\pi}{3M_4^2}(\rho_{\phi}+\rho)(1+\frac{\rho_{\phi}+\rho}{2\lambda})
\end{equation}
Cosmological perturbations of warm inflation model (with viscous pressure) have been studied in Ref.\cite{9-f} (\cite{2-ne}).
Warm tachyon inflationary universe model (on the brane) have been studied in
Ref.\cite{1-m} (\cite{v-1}). Warm inflation on the brane (with viscous pressure) have been studied in Ref \cite{6-f}.  As far as, we know, a model in which warm tachyon inflation with viscous pressure on the brane
has not been yet considered.
Therefor, in the present work we will study warm-tachyon
inspired inflation with viscous pressure on the brane, using
the above modified Friedmann
equation. The paper organized as: In the next section we will
describe warm-tachyon inflationary universe model with viscous pressure on the brane. In section (3) we obtain the perturbation parameters
for our model. In section (4) we study our model  using the
exponential potential in high dissipative regime and high energy limit for two cases: 1- Dissipative parameter $\Gamma$ and bulk viscous parameter $\zeta$ are constant parameters.
2- Dissipative parameter  as a function of tachyon field $\phi$ and bulk viscous parameter
as a function of radiation-matter mixture  energy density $\rho$. Finally in
section (5) we close by some concluding  remarks.

\section{The model}
In this section, we will obtain the characteristics of warm tachyon inflation with viscous pressure on the brane in the background level. This model may be described  by an effective tachyon fluid and matter-radiation imperfect fluid. The energy-momentum tensor of tachyon fluid in a spatially flat Friedmann Robertson Walker
(FRW) is recognized by
$T_{\mu}^{\nu}=diag(-\rho_{\phi},P_{\phi},P_{\phi},P_{\phi})$.
 The pressure and energy density of tachyon field are defined by \cite{1}

 \hspace{1.5mm}

\begin{eqnarray}\label{2}
P_{\phi}=-V(\phi)\sqrt{1-\dot{\phi}^2}
\end{eqnarray}

and
\begin{eqnarray}\label{3}
\rho_{\phi}=\frac{V(\phi)}{\sqrt{1-\dot{\phi}^2}}~~~~~~~
\end{eqnarray}

respectively, where $V(\phi)$ is the effective scalar potential
associated with tachyon field $\phi$. Important characteristics
of this potential are $\frac{dV}{d\phi}<0$ and $V(\phi\rightarrow
 \infty)\rightarrow 0$ \cite{2}. The imperfect fluid is a mixture of matter and radiation of adiabatic index $\gamma$ which have  energy density $\rho=Ts(\phi,T)$ ($T$ is the temperature  and $s$ is the entropy density of the imperfect fluid.) and pressure $P+\Pi$, where $P=(\gamma-1)\rho$. Also,  $\Pi=-3\zeta H$ is bulk viscous pressure  \cite{3-ne}, where $H$ is
Hubble parameter and $\zeta$ is phenomenological coefficient of bulk viscosity.
This coefficient is positive-definite by the second law of thermodynamics and in general depends on the energy density $\rho$ of the fluid.
The dynamic of warm tachyon
inflation with viscous pressure on the brane  in spatially flat FRW model  is described by Friedmann equation,
\begin{eqnarray}\label{4}
3H^2=(\frac{V(\phi)}{\sqrt{1-\dot{\phi}^2}}+\rho)(1+\frac{1}{2\lambda}[\frac{V(\phi)}{\sqrt{1-\dot{\phi}^2}}+\rho])
\end{eqnarray}

conservation equations of tachyon field  and imperfect fluid which are connected by the dissipation term $\Gamma\dot{\phi}^2$
\begin{equation}\label{5}
\dot{\rho}_{\phi}+3H(P_{\phi}+\rho_{\phi})=-\Gamma\dot{\phi}^2\Rightarrow
\frac{\ddot{\phi}}{1-\dot{\phi}^2}+3H\dot{\phi}+\frac{V'}{V}=-\frac{\Gamma}{V}\sqrt{1-\dot{\phi}^2}\dot{\phi}
\end{equation}

and
\begin{equation}\label{6}
\dot{\rho}+3H(\rho+P+\Pi)=\dot{\rho}+3H(\gamma\rho+\Pi)=\Gamma\dot{\phi}^2
\end{equation}

where
$\Gamma$ is the dissipative coefficient with the dimension
$M_4^5$
\hspace{1.5mm}. In the above equations dots "." mean derivative with
respect to cosmic time, prime  denotes derivative with respect
to tachyon field $\phi$ and in Eq.(\ref{4}) we choose $\frac{8\pi}{M_4^2}=1$.
We would like to express the evolution equation (\ref{6})  in terms of entropy
density $s(\phi,T)$. This parameter is defined by a thermodynamical relation \cite{nn-1}
\begin{equation}\label{}
s(\phi,T)=-\frac{\partial f}{\partial T}=-\frac{\partial V}{\partial T}
\end{equation}
in terms of the Helmholtz free energy $f$
\begin{equation}\label{}
f=\rho_{T}-Ts=\frac{V(\phi,T)}{\sqrt{1-\dot{\phi}^2}}+\rho-Ts
\end{equation}
which  is dominated by the thermodynamical potential $V(\phi, T)$ in slow-roll limit. The total energy density and total pressure are given by
\begin{eqnarray}\label{}
\rho_T=\frac{V}{\sqrt{1-\dot{\phi}^2}}+Ts~~~~~~~~~~~~~~~~\\
\nonumber
P_T=-\frac{V}{\sqrt{1-\dot{\phi}^2}}+(\gamma-1)Ts+\Pi
\end{eqnarray}
The viscous pressure for an expanding universe is negative ($\Pi=-3\zeta H$), therefore this term acts to decrease the total pressure. Using Eq.(\ref{6}), we can find the entropy density evolution for our model as
\begin{eqnarray}\label{nn-1}
T\dot{s}+3H(\gamma Ts+\Pi)=\Gamma\dot{\phi}^2
\end{eqnarray}
where for  a quasi-equilibrium high temperature thermal bath as an inflation fluid, we have $\gamma=\frac{4}{3}$. The bulk viscosity effects may be read from above equation. Thus bulk viscous pressure $\Pi$ as a negative quantity, enhances the source of entropy density on the RHS of the  evolution equation (\ref{nn-1}). Therefore,  energy density of radiation  and entropy density increase by the bulk viscosity pressure $\Pi$ (see FIG.1 and FIG.2).
During inflation epoch the energy density
(\ref{3}) is the order of potential $\rho_{\phi}\sim V$ and
dominates over the imperfect fluid energy density $\rho_{\phi}>\rho$, this limit is called stable regime \cite{nn-1}.
Using slow-roll approximation when $\dot{\phi}\ll 1$ and
$\ddot{\phi}\ll(3H+\frac{\Gamma}{V})$ \cite{3}  the
dynamic equations (\ref{4}) and (\ref{5})   are reduced to
\begin{equation}\label{7}
3H^2=V(1+\frac{V}{2\lambda})
\end{equation}

and
\begin{equation}\label{8}
3H(1+r)\dot{\phi}=-\frac{V'}{V}
\end{equation}

where $r=\frac{\Gamma}{3HV}$.\hspace{1.5mm}
From above equations and
Eq.(\ref{6}),  when the decay of the tachyon field to imperfect fluid is quasi-stable i.e.  $\dot{\rho}\ll 3H(\gamma\rho+\Pi)$ and $\dot{\rho}\ll\Gamma\dot{\phi}^2$, $\rho$ could be written as
\begin{equation}\label{9}
\rho=\frac{1}{\gamma}(\frac{\Gamma}{3H}\dot{\phi}^2-\Pi)=\frac{1}{\gamma}(\frac{r}{3(1+\frac{V}{2\lambda})(1+r)^2}(\frac{V'}{V})^2-\Pi)
\end{equation}

In the present work we will restrict our analysis in high dissipative regime i.e. $r\gg 1$ where the dissipation coefficient $\Gamma$ is much greater than $3HV$. The reason of this choice is the following. In weak dissipative i.e. $r\ll 1$, the expansion of the universe in the inflation era disperses the decay of the inflaton, so there is a little  chance for interaction between the sectors of the inflationary fluid and we do not have non-negligible bulk viscosity. Warm inflation in high and weak dissipative regimes for a model without bulk viscous pressure have been studied in Refs. \cite{3} and \cite{1-ne}  respectively. Dissipation parameter $\Gamma$ may be a positive function of inflaton $\phi$ and temperature $T$ or a constant parameter by the second law of thermodynamics.
There are some specific forms for the dissipative coefficient, with the most common which are found in the literatures being the $\Gamma\sim T^3$ form \cite{nn-1},\cite{2nn},\cite{3nn},\cite{4nn}.
We like quasi-stable condition in the inflation epoch \cite{nn-1} where the energy density of radiation (which is obtained from dissipation of inflaton field) is smaller than the energy of tachyon field $\rho<\rho_{\phi}\sim V(\phi)$. In warm-tachyon model of inflation the interesting potential is descending potential in term of tachyon field $\phi$ ($V(\phi)=V_0\exp(-\alpha\phi)$) \cite{1-m}. Therefore we need a descending form of dissipation in term of tachyon field $\phi$. An important choice for dissipation parameter $\Gamma$ as a function of tachyon field is a descending function $\Gamma(\phi)\propto \exp(-\alpha\phi)$ \cite{1-m},\cite{2-ne}. In some works parameter $\Gamma$ and potential of the inflaton have the same form \cite{1-m}. In Ref.\cite{2-ne}, perturbation parameters for warm inflationary model with viscous pressure have been obtained where $\Gamma=\Gamma(\phi)=\alpha_1V(\phi)$ and $\Gamma=\Gamma_0=const$. In this work we will study the warm-tachyon inflationary universe  model with viscous pressure on the brane for these two cases. \\
We introduce the slow-roll parameters
for our model as
\begin{eqnarray}\label{10}
\epsilon=-\frac{\dot{H}}{H^2}\simeq\frac{1}{2(1+r)V}(\frac{V'}{V})^2\frac{1+\frac{V}{\lambda}}{(1+\frac{V}{2\lambda})^2}
\end{eqnarray}

and
\begin{eqnarray}\label{11}
\eta=-\frac{\ddot{H}}{H\dot{H}}\simeq\frac{V'}{V^2(1+r)[1+\frac{V}{2\lambda}]}~~~~~~~~~~~~~~~~~~~~\\
\nonumber \times
[\frac{2V''}{V'}-\frac{V'}{V}-\frac{r'}{(1+r)}+\frac{V'}{\lambda+V}]-2\epsilon
\end{eqnarray}

A relation between two energy densities $\rho_{\phi}$ and
$\rho$ is obtained from Eqs. (\ref{9}) and (\ref{10})
\begin{equation}\label{12}
\rho=\frac{1}{\gamma}(\frac{2}{3}\frac{r(1+\frac{V}{2\lambda})}{(1+r)(1+\frac{V}{\lambda})}\epsilon\rho_{\phi}-\Pi)
\end{equation}

The condition of inflation epoch $\ddot{a}>1$ could be obtained
by inequality $\epsilon<1$. Therefore from above equation,
warm-tachyon inflation with viscous pressure
could take place when
\begin{equation}\label{13}
\frac{1+\frac{\rho_{\phi}}{2\lambda}}{1+\frac{\rho_{\phi}}{\lambda}}\rho_{\phi}>\frac{3(1+r)}{2r}[\gamma\rho+\Pi]
\end{equation}

Inflation period ends when $\epsilon\simeq 1$ which implies
\begin{equation}\label{14}
\frac{1+\frac{\rho_{\phi}}{2\lambda}}{1+\frac{\rho_{\phi}}{\lambda}}\rho_{\phi}\simeq\frac{3(1+r)}{2r}[\gamma\rho+\Pi]
\end{equation}

or equivalently
\begin{equation}\label{}
\nonumber
\frac{1+\frac{V_f}{\lambda}}{1+\frac{V_f}{2\lambda}}[\frac{V'_f}{V_f}]^2\frac{1}{V_f}\simeq 2(1+r_f)
\end{equation}

where the subscript $f$ denotes the end of inflation. The number
of e-folds is given by
\begin{eqnarray}\label{15}
N=\int_{\phi_{*}}^{\phi_f}Hdt=\int_{\phi_{*}}^{\phi_f}\frac{H}{\dot{\phi}}d\phi=-\int_{\phi_{*}}^{\phi_f}\frac{V^2}{V'}(1+r)(1+\frac{V}{2\lambda})d\phi
\end{eqnarray}

where the subscript $*$ denotes the epoch when the cosmological
scale exits the horizon.
\section{Perturbation}
In this section we will study inhomogeneous perturbations of the FRW background. These perturbations in the longitudinal gauge, may be described by the perturbed FRW metric
\begin{equation}\label{16}
ds^2=(1+2\Phi)dt^2-a^2(t)(1-2\Psi)\delta_{ij}dx^idx^j
\end{equation}

where $\Phi$ and $\Psi$ are gauge-invariant metric perturbation variables \cite{7-f}. All perturbed quantities have a spatial sector of the form $e^{i\mathbf{kx}}$, where $k$ is the wave number. Perturbed Einstein field equations in momentum space have only the temporal parts
\begin{equation}\label{}
\nonumber
\Phi=\Psi
\end{equation}

\begin{equation}\label{17}
\dot{\Phi}+H\Phi=\frac{1}{2}[-\frac{(\gamma\rho+\Pi)av}{k}+\frac{V\dot{\phi}}{\sqrt{1-\dot{\phi}^2}}\delta\phi](1+\frac{V}{\lambda})
\end{equation}

\begin{eqnarray}\label{18}
\frac{\ddot{\delta\phi}}{1-\dot{\phi}^2}+[3H+\frac{\Gamma}{V}]\dot{\delta\phi}+[\frac{k^2}{a^2}+(\ln V)''+\dot{\phi}(\frac{\Gamma}{V})']\delta\phi\\
\nonumber
-[\frac{1}{1-\dot{\phi}^2}+3]\dot{\phi}\dot{\Phi}-[\dot{\phi}\frac{\Gamma}{V}-2(\ln V)']\Phi=0
\end{eqnarray}

\begin{eqnarray}\label{19}
(\dot{\delta\rho})+3\gamma H\delta\rho+ka(\gamma\rho+\Pi)v+3(\gamma\rho+\Pi)\dot{\Phi}\\
\nonumber
-\dot{\phi}^2\Gamma'\delta\phi-\Gamma\dot{\phi}[2(\dot{\delta\phi})+\dot{\phi}\Phi]=0~~~~~~~~~~~~~~
\end{eqnarray}
and
\begin{equation}\label{20}
\dot{v}+4Hv+\frac{k}{a}[\Phi+\frac{\delta P}{\rho+P}+\frac{\Gamma\dot{\phi}}{\rho+P}\delta\phi]=0
\end{equation}
where
\begin{equation}\label{}
\nonumber
\delta P=(\gamma-1)\delta\rho+\delta\Pi~~~~~~~\delta\Pi=\Pi[\frac{\zeta_{,\rho}}{\zeta}\delta\rho+\Phi+\frac{\dot{\Phi}}{H}]
\end{equation}

The above equations are obtained for Fourier components $e^{i\mathbf{kx}}$, where the subscript $k$ is omitted. $v$ in the above set of equations is found from the decomposition of the velocity field ($\delta u_j=-\frac{iak_J}{k}ve^{i\mathbf{kx}}, j=1,2,3$) \cite{6-f}. Warm inflation model may be considered as a hybrid-like inflationary model where the inflaton field interacts with radiation field \cite{9-f}, \cite{8-f}. Entropy perturbation may be related to dissipation term \cite{10-f}. During inflationary phase with slow-roll approximation, for non-decreasing adiabatic modes on large scale limit $k\ll aH$, we assume that the perturbed quantities could not vary strongly. So we have $H\Phi\gg\dot{\Phi}$, $(\ddot{\delta\phi})\ll(\Gamma+3H)(\dot{\delta\phi})$, $(\dot{\delta\rho})\ll\delta\rho$ and $\dot{v}\ll 4Hv$. In slow-roll limit and by using the above limitations, the set of perturbed equations are reduced to
\begin{equation}\label{21}
\Phi\simeq \frac{1}{2H}[-\frac{4(\gamma\rho+\Pi)av}{k}+V\dot{\phi}\delta\phi][1+\frac{V}{\lambda}]
\end{equation}

\begin{equation}\label{22}
[3H+\frac{\Gamma}{V}]\dot{\delta\phi}+[(\ln V)''+\dot{\phi}(\frac{\Gamma}{V})']\delta\phi
\simeq[\dot{\phi}\frac{\Gamma}{V}-2(\ln V)']\Phi
\end{equation}

\begin{equation}\label{23}
\delta\rho\simeq\frac{\dot{\phi}^2}{3\gamma H}[\Gamma'\delta\phi+\Gamma\Phi]
\end{equation}
and
\begin{eqnarray}\label{24}
v\simeq-\frac{k}{4aH}(\Phi+\frac{(\gamma-1)\delta\rho+\delta\Pi}{\gamma\rho+\Pi}+\frac{\Gamma\dot{\phi}}{\gamma\rho+\Pi}\delta\phi)
\end{eqnarray}
Using Eqs.(\ref{21}), (\ref{23}) and (\ref{24}) perturbation variable $\Phi$ is determined
\begin{eqnarray}\label{25}
\Phi\simeq\frac{\dot{\phi}V}{2H}\frac{\delta\phi}{G(\phi)}[1+\frac{\Gamma}{4HV}+([\gamma-1]+\Pi\frac{\zeta_{,\rho}}{\zeta})\frac{\dot{\phi}\Gamma'}{12\gamma H^2V}][1+\frac{V}{\lambda}]
\end{eqnarray}
where
\begin{eqnarray}\label{}
\nonumber
G(\phi)=1-\frac{1}{8H^2}[2\gamma\rho+3\Pi+\frac{\gamma\rho+\Pi}{\gamma}(\Pi\frac{\zeta_{,\rho}}{\zeta}-1)][1+\frac{V}{\lambda}]
\end{eqnarray}

In the above equations, for $\Pi\rightarrow 0$ and $\gamma=\frac{4}{3}$ case, we obtain warm tachyon inflation model without viscous pressure on the brane \cite{v-1} (In this case, we find $G(\phi)\rightarrow 1$ because of the inequality  $\frac{\rho}{V}\ll 1$.).

We can solve the above equations by taking tachyon field $\phi$ as the independent variable in place of cosmic time $t$. Using Eq.(\ref{8}) we find
\begin{eqnarray}\label{26}
(3H+\frac{\Gamma}{V})\frac{d}{dt}=(3H+\frac{\Gamma}{V})\dot{\phi}\frac{d}{d\phi}=-\frac{V'}{V}\frac{d}{d\phi}
\end{eqnarray}

From above equation, Eq.(\ref{22}) and Eq.(\ref{25}), the expression $\frac{(\delta\phi)'}{\delta\phi}$ is obtained
\begin{eqnarray}\label{27}
\frac{(\delta\phi)'}{\delta\phi}=\frac{1}{(\ln V)'}[(\ln V)''+\dot{\phi}(\frac{\Gamma}{V})'+(2(\ln V)'-\dot{\phi}\frac{\Gamma}{V})(\frac{V\dot{\phi}}{2GH})\\
\nonumber
\times (1+\frac{\Gamma}{4HV}+[(\gamma-1)+\Pi\frac{\zeta_{,\rho}}{\zeta}]\frac{\dot{\phi}\Gamma'}{12\gamma H^2 V})(1+\frac{V}{\lambda})]~~~~~~~~~~~~~~
\end{eqnarray}

We will return to the above relation soon. Following Refs.\cite{1-m}, \cite{6-f} and  \cite{10-f},  we introduce auxiliary function $\chi$ as
\begin{equation}\label{28}
\chi=\frac{\delta\phi}{(\ln V)'}\exp[\int\frac{1}{3H+\frac{\Gamma}{V}}(\frac{\Gamma}{V})'d\phi]
\end{equation}

From above definition we have
\begin{eqnarray}\label{29}
\frac{\chi'}{\chi}=\frac{(\delta\phi)'}{\delta\phi}-\frac{(\ln V)''}{(\ln V)'}+\frac{(\frac{\Gamma}{V})'}{3H+\frac{\Gamma}{V}}
\end{eqnarray}

Using above equation and Eq.(\ref{27}) we find
\begin{eqnarray}\label{30}
\frac{\chi'}{\chi}=(-\frac{\dot{\phi}}{(\ln V)'}\frac{\Gamma}{V}+2)
(\frac{V\dot{\phi}}{2H})~~~~~~~~~~~~~~~~~~~~~~~~~~~~~~~~~~~\\
\nonumber
\times[1+\frac{\Gamma}{4HV}+[(\gamma-1)+\Pi\frac{\zeta_{,\rho}}{\zeta}]\frac{\dot{\phi}\Gamma'}{12\gamma H^2 V}][1+\frac{V}{\lambda}]
\end{eqnarray}

We could rewrite this equation, using Eqs. (\ref{7}) and (\ref{8})
\begin{eqnarray}\label{31}
\frac{\chi'}{\chi}=-\frac{9}{8G}\frac{2H+\frac{\Gamma}{V}}{(3H+\frac{\Gamma}{V})^2}[\Gamma+4HV-([\gamma-1]+\Pi\frac{\zeta_{,\rho}}{\zeta})
\\
\nonumber
\times\frac{\Gamma'(\ln V)'}{3\gamma H(3H+\frac{\Gamma}{V})}]\frac{(\ln V)'}{V}\frac{1+\frac{V}{\lambda}}{1+\frac{V}{2\lambda}}~~~~~~~~
\end{eqnarray}

A solution for the above equation is
\begin{eqnarray}\label{32}
\chi(\phi)=C\exp(-\int\{-\frac{9}{8G}\frac{2H+\frac{\Gamma}{V}}{(3H+\frac{\Gamma}{V})^2} [\Gamma+4HV-([\gamma-1]+\Pi\frac{\zeta_{,\rho}}{\zeta})\\
\nonumber
\times\frac{\Gamma'(\ln V)'}{3\gamma H(3H+\frac{\Gamma}{V})}]\frac{1+\frac{V}{\lambda}}{1+\frac{V}{2\lambda}}\frac{(\ln V)'}{V}\}d\phi)
\end{eqnarray}

where $C$ is integration constant. From above equation and Eq.(\ref{29}) we find small change of variable $\delta\phi$
\begin{eqnarray}\label{33}
\delta\phi=C(\ln V)'\exp(\Im(\phi))
\end{eqnarray}
where
\begin{eqnarray}\label{34}
\Im(\phi)=-\int[\frac{(\frac{\Gamma}{V})'}{3H+\frac{\Gamma}{V}}+(\frac{9}{8G}\frac{2H+\frac{\Gamma}{V}}{(3H+\frac{\Gamma}{V})^2}~~~~~~~~~~~~~~~~~~~~~~~~~~~~~~~\\
\nonumber
\times[\Gamma+4HV-([\gamma-1]+\Pi\frac{\zeta_{,\rho}}{\zeta})\frac{\Gamma'(\ln V)'}{3\gamma H(3H+\frac{\Gamma}{V})}]\frac{1+\frac{V}{\lambda}}{1+\frac{V}{2\lambda}}\frac{(\ln V)'}{V}]d\phi
\end{eqnarray}

 Perturbed matter fields of our model are inflaton $\delta\phi$, radiation $\delta\rho$ and velocity $k^{-1}(P+\rho)v_{,i}$. We can explain the cosmological perturbations in terms of gauge-invariant variables. These variables are important for development of perturbation after the end of inflation period. The curvature perturbation $\mathfrak{R}$ and entropy perturbation $e$ are defied by \cite{nn-2,nn-3}
\begin{eqnarray}\label{}
\mathfrak{R}=\Phi-k^{-1}aHv~~~~~~~\\
\nonumber
e=\delta P-c_s^2\delta\rho~~~~~~
\end{eqnarray}
where $c_s^2=\frac{\dot{P}}{\dot{\rho}}$, and
\begin{eqnarray}\label{}
\Phi\simeq C\frac{\dot{\phi}V}{2HG(\phi)}[1+\frac{\Gamma}{4HV}+([\gamma-1]+\Pi\frac{\zeta_{,\rho}}{\zeta})\frac{\dot{\phi}\Gamma'}{12\gamma H^2V}][1+\frac{V}{\lambda}] (\ln V)'\exp(\Im(\phi))
\end{eqnarray}
 In large scale limit, where $k\ll aH$, and with slow roll condition, the curvature perturbation is given by
\begin{equation}\label{}
 \mathfrak{R}\sim C
\end{equation}
and the entropy perturbation vanishes \cite{nn-3}.
We can find the density perturbation amplitude by using the above equation and Eqs.(12), (\ref{33})  \cite{12-f}
\begin{eqnarray}\label{35}
P_R^{\frac{1}{2}}\sim \mathfrak{R}\sim C~~~~~~~~~~~~~~~~~~~~~~~~~~~~~~~~~~~~~~~~~~~~~~~\\
\nonumber
\delta_H=\frac{2}{5}P_R^{\frac{1}{2}}=\frac{16\pi}{5}\frac{\exp(-\Im(\phi))}{(\ln V)'}\delta\phi=\frac{16\pi}{15}\frac{\exp(-\Im(\phi))}{Hr\dot{\phi}}\delta\phi
\end{eqnarray}

For high or low energy limit ($V\gg\lambda$ or $V\ll \lambda$) and by inserting $\Gamma=0$, the above equation reduces to $\delta_{H}\simeq\frac{H}{\dot{\phi}}\delta\phi$ which agrees with the density perturbation in cool inflation model \cite{1-i}. In warm inflation model the fluctuations of the scalar field in high dissipative regime ($r\gg 1$) may be generated by thermal fluctuation instead of quantum fluctuations \cite{5} as
\begin{equation}\label{36}
(\delta\phi)^2\simeq\frac{k_F T_r}{2\pi^2}
\end{equation}

where in this limit freeze-out wave number $k_F=\sqrt{\frac{\Gamma H}{V}}=H\sqrt{3r}\geq H$ corresponds to the freeze-out scale at the point when, dissipation damps out to thermally excited fluctuations ($\frac{V''}{V'}<\frac{\Gamma H}{V}$) \cite{5}. With the help of the above equation and Eq.(\ref{35}) in high dissipative regime ($r\gg 1$) and high energy limit ($\lambda\gg 1$) we find
\begin{equation}\label{37}
\delta_H^2=\frac{128\sqrt{3}}{75}\frac{\exp(-2\tilde{\Im}(\phi))}{\sqrt{r}\tilde{\epsilon}}\frac{T_r}{H}
\end{equation}

where
\begin{equation}\label{38}
\tilde{\Im}(\phi)=-\int[\frac{1}{3Hr}(\frac{\Gamma}{V})'+\frac{9}{4\tilde{G}}(1-[(\gamma-1)+\Pi\frac{\zeta_{,\rho}}{\zeta}]\frac{(\ln\Gamma)'(\ln V)'}{9\gamma rH^2})(\ln V)']d\phi
\end{equation}

\begin{equation}\label{}
\nonumber
\tilde{G}(\phi)=1-\frac{V}{8\lambda H^2}[2\gamma\rho+3\Pi+\frac{\gamma\rho+\Pi}{\gamma}(\Pi\frac{\zeta_{,\rho}}{\zeta}-1)]
\end{equation}

and
\begin{equation}\label{39}
\tilde{\epsilon}=\frac{2\lambda}{rV^2}(\frac{V'}{V})^2
\end{equation}

An important perturbation parameter is scalar index $n_s$ which
in high dissipative regime is given by
\begin{equation}\label{40}
n_s=1+\frac{d\ln \delta_H^2}{d\ln k}\approx
1-4\tilde{\eta}+\tilde{\epsilon}-(\frac{8\lambda\tilde{\epsilon}}{rV^2})^{\frac{1}{2}}(\tilde{\Im}'(\phi)+\frac{r'}{4r})
\end{equation}

where
\begin{equation}\label{41}
\tilde{\eta}=\frac{M_4^2\lambda}{4\pi r}\frac{V'}{V^3}[\frac{2V''}{V'}-\frac{r'}{r}]-2\tilde{\epsilon}
\end{equation}

In Eq.(\ref{40}) we have used a relation between small change of
the number of e-folds and interval in wave number ($dN=-d\ln k$).
Running of the scalar spectral index may be found as
\begin{eqnarray}\label{42}
\alpha_s=\frac{dn_s}{d\ln k}=-\frac{dn_s}{dN}=-\frac{d\phi}{dN}\frac{dn_s}{d\phi}=\frac{2\lambda V'}{rV^3}n_s'
\end{eqnarray}

This parameter is one of the interesting cosmological
perturbation parameters which is approximately $-0.038$, by using
WMAP7 observational results \cite{6}.\\ During inflation epoch,
there are two independent components of gravitational waves
($h_{\times +}$) with action of massless scalar field are
produced by the generation of tensor perturbations. The amplitude
of tensor perturbation is given by
\begin{eqnarray}\label{43}
A_g^2=2(\frac{H}{2\pi})^2\coth[\frac{k}{2T}]=\frac{V^2}{12\lambda\pi^2}\coth[\frac{k}{2T}]
\end{eqnarray}

where, the temperature $T$ in extra factor $\coth[\frac{k}{2T}]$
denotes, the temperature of the thermal background of
gravitational wave \cite{7}. Spectral index $n_g$ may be found as
\begin{eqnarray}\label{44}
n_g=\frac{d}{d\ln k}(\ln [\frac{A_g^2}{\coth(\frac{k}{2T})}])\simeq-2\tilde{\epsilon}
\end{eqnarray}
where $A_g\propto k^{n_g}\coth[\frac{k}{2T}]$ \cite{7}.  Using Eqs. (\ref{37})  and
(\ref{43}) we write the tensor-scalar ratio in high dissipative
regime
\begin{eqnarray}\label{45}
R(k_0)=\frac{A_g^2}{P_R}|_{k=k_{0}}=\frac{\sqrt{3}}{64\pi^2}\frac{r^{\frac{1}{2}}\tilde{\epsilon}H^3}{T_r}\exp(2\Im(\phi))\coth[\frac{k}{2T}]|_{k=k_{0}}
\end{eqnarray}

where $k_{0}$ is referred  to pivot point \cite{7} and $P_R=\frac{25}{4}\delta_H^2$. An upper bound for this parameter is obtained
by using    WMAP7 data, $R<0.36$ \cite{6}.

\section{Exponential potential }

In this section we consider our model with the tachyonic
effective potential
\begin{equation}\label{46}
V(\phi)=V_0\exp(-\alpha\phi)
\end{equation}

where parameter $\alpha>0$  is related to mass
of the tachyon field \cite{8}. The exponential form of potential have
characteristics of tachyon field ($\frac{dV}{d\phi}<0$ and
$V(\phi\rightarrow 0)\rightarrow V_{max}$ ). We develop our model
in high dissipative regime i.e. $r\gg 1$ and high energy limit i.e. $V(\phi)\gg \lambda$, for two cases: 1- $\Gamma$ and $\zeta$ are constant parameters. 2- $\Gamma$ as a function of tachyon  field $\phi$ and $\zeta$ as a function of energy density  $\rho$ of imperfect fluid.

\subsection{$\Gamma=\Gamma_0$, $\zeta=\zeta_0$ case}
From Eq.(\ref{39}) slow-roll parameter $\tilde{\epsilon}$ in the present case has the form
\begin{equation}\label{47}
\tilde{\epsilon}=\frac{2\lambda\alpha^2}{rV_0^2}\exp(2\alpha\phi)
\end{equation}


We find the evolution of tachyon field with the help of Eq.(\ref{8})
\begin{equation}\label{50}
\phi(t)=\frac{1}{\alpha}\ln[\frac{\alpha^2 V_0}{\Gamma_0}t+e^{\alpha\phi_i}]
\end{equation}

where $\phi_i=\phi(t=0)$.

Dissipation parameter $r=\frac{\Gamma}{3HV}$ in this case is given by
\begin{equation}\label{49}
r=\frac{\sqrt{2\lambda}\Gamma_0}{\sqrt{3}V_0^2}\exp(2\alpha\phi)\gg 1
\end{equation}

Hubble parameter for our model has the form
\begin{equation}\label{51}
H=\frac{V_0}{\sqrt{6\lambda}}\exp(-\alpha\phi)
\end{equation}



Using Eqs.(\ref{12}) and (\ref{47}), the energy density of the radiation-matter fluid in high dissipative limit becomes
\begin{equation}\label{54}
\rho=\frac{V_0\exp(-\alpha\phi)}{\gamma}(\frac{4}{3}\frac{\sqrt{6\lambda}}{\Gamma_0}\alpha^2+\frac{3\zeta_0}{\sqrt{6\lambda}})
\end{equation}

and, in terms of tachyon field energy density $\rho_{\phi}$ becomes
\begin{equation}\label{55}
\rho=\frac{\rho_{\phi}}{\gamma}(\frac{4}{3}\frac{\sqrt{6\lambda}}{\Gamma_0}\alpha^2+\frac{3\zeta_0}{\sqrt{6\lambda}})
\end{equation}
For this example, the  entropy density in terms of cosmic time may be obtained from Eqs. (\ref{50}),(\ref{54})
\begin{equation}\label{}
Ts=(\frac{4}{3}\frac{\sqrt{6\lambda}}{\Gamma_0}\alpha^2+\frac{3\zeta_0}{\sqrt{6\lambda}})\frac{1}{(e^{\alpha\phi_i}+\frac{\alpha^2 V_0}{\Gamma_0}t)\gamma}
\end{equation}
In FIG.1,  we plot the entropy density in terms of cosmic time. As one can see the entropy density increases by the bulk viscous effect \cite{nn-1}.

\begin{figure}[h]
\centering
  \includegraphics[width=10cm]{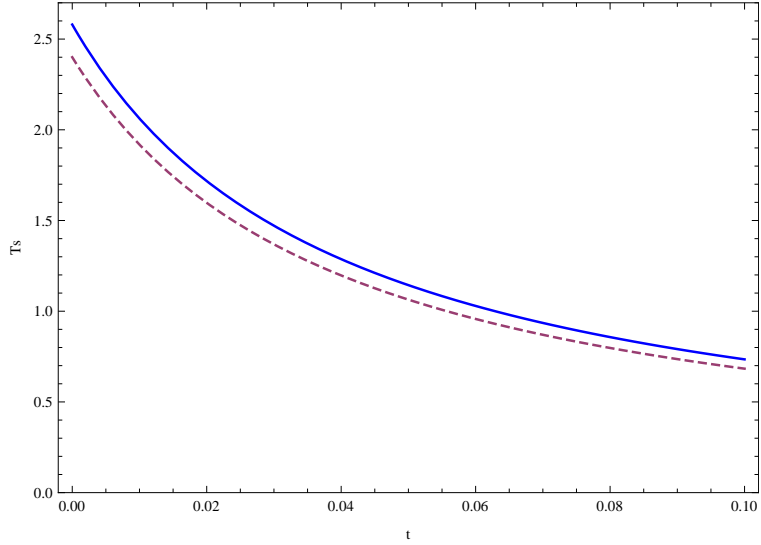}
  \caption{We plot the entropy density $s$ in terms of cosmic time $t$ where, $\Pi=0$ by dashed curve and $\Pi=-3\zeta_0 H$ by blue curve ($T=5.47\times 10^{-5}$, $\Gamma=\Gamma_0=8\pi\times 10^{-3}, \zeta_0=(8\pi)^3\times10^{-6}, \alpha=+8\pi\times10^{-2}$)}
 \label{fig:F1}
\end{figure}
From Eq.(\ref{15}), the number of e-fold at the end of inflation, by using the potential (\ref{46}), for our inflation model is given by

\begin{equation}\label{57}
N_{total}=\frac{\sqrt{6\lambda}\Gamma_0}{\alpha}(\phi_f-\phi_i)
\end{equation}

where $\phi_f>\phi_i$. Using Eqs.(\ref{37}) and (\ref{45}), we could find the scalar spectrum and scalar-tensor ratio
\begin{equation}\label{58}
\delta_H^2=\frac{128}{25}\frac{\Gamma_0^{\frac{3}{2}}T_r}{\alpha^2(6\lambda)^{\frac{1}{4}}}V(\phi_0)^{\frac{9}{2\tilde{G}}-2}
\end{equation}
where
\begin{eqnarray}\label{}
\nonumber
\tilde{G}=1-\frac{1}{4}[2\gamma a_1-\frac{9\zeta_0}{\sqrt{6\lambda}}-\frac{1}{\gamma}(\gamma a_1-\frac{3\zeta_0}{\sqrt{6\lambda}})]\\
\nonumber
a_1=\frac{1}{\gamma}(\frac{4}{3}\frac{\sqrt{6\lambda}}{\Gamma_0}\alpha^2+\frac{3\zeta_0}{\sqrt{6\lambda}})
\end{eqnarray}

and
\begin{equation}\label{59}
R=\frac{1}{64\pi^2}\frac{\alpha^2}{\Gamma_0^{\frac{3}{2}}T_r(6\lambda)^{\frac{1}{4}}}V(\phi_0)^{4-\frac{9}{2\tilde{G}}}\coth[\frac{k}{2T}]\mid_{k=k_0}
\end{equation}
In the above equation we have used the Eq.(\ref{38}) where
\begin{equation}\label{}
\tilde{\Im}(\phi)=\ln(\frac{V^{1-\frac{9}{4\tilde{G}}}}{\Gamma_0})
\end{equation}

These parameters may by restricted by WMAP7 observational data \cite{6}.
\subsection{$\Gamma=\Gamma(\phi)$, $\zeta=\zeta(\rho)$ case}
Now we assume $\zeta=\zeta(\rho)=\zeta_1\rho$ and $\Gamma=\Gamma(\phi)=\alpha_1 V(\phi)=\alpha_1 V_0\exp(-\alpha\phi)$ (see appendix B), where $\alpha_1$ and $\zeta_1$ are positive constants. By using exponential potential (\ref{46}), Hubble parameter, $r$ parameter and slow-roll parameter $\tilde{\epsilon}$  have the form
\begin{eqnarray}\label{60}
H(\phi)=\frac{V_0}{\sqrt{6\lambda}}\exp(-\alpha\phi)~~~~~~~~~~~~r=\frac{\alpha_1\sqrt{2\lambda}}{\sqrt{3}V_0}\exp(\alpha\phi)~~~~~~\\
\nonumber
\tilde{\epsilon}=\frac{2\lambda\alpha^2}{rV_0^2}\exp(2\alpha\phi)~~~~~~~~~~~~~~~~~~~~~~~~~~~~~~~~~~~~~~~~~~
\end{eqnarray}

respectively. Using Eq.(\ref{8}), we find the scalar field $\phi$ in term of cosmic time
\begin{equation}\label{61}
\phi(t)=-\frac{\alpha}{\alpha_1}t+\phi_i
\end{equation}

The energy density of imperfect fluid $\rho$ in term of the inflaton energy density $\rho_{\phi}$ is given by the expression
\begin{equation}\label{62}
\rho=\frac{(6\lambda)^{\frac{1}{2}}\alpha^2}{3\alpha_1\gamma}[1+\frac{\sqrt{3}\zeta_1}{\gamma\sqrt{2\lambda}}\rho_{\phi}]^{-1}
\end{equation}
We can find the entropy density $s$ in terms of cosmic time
\begin{equation}\label{}
Ts=\frac{(6\lambda)^{\frac{1}{2}}\alpha^2}{3\alpha_1\gamma}[1+\frac{\sqrt{3}\zeta_1 V_0}{\gamma\sqrt{6\lambda}}\exp(\frac{-\alpha^2}{\alpha_1}t+\phi_i t)]^{-1}
\end{equation}

The entropy density and energy density of our model in this case increase by the bulk viscosity effect (see FIG.2).
\begin{figure}[h]
\centering
  \includegraphics[width=10cm]{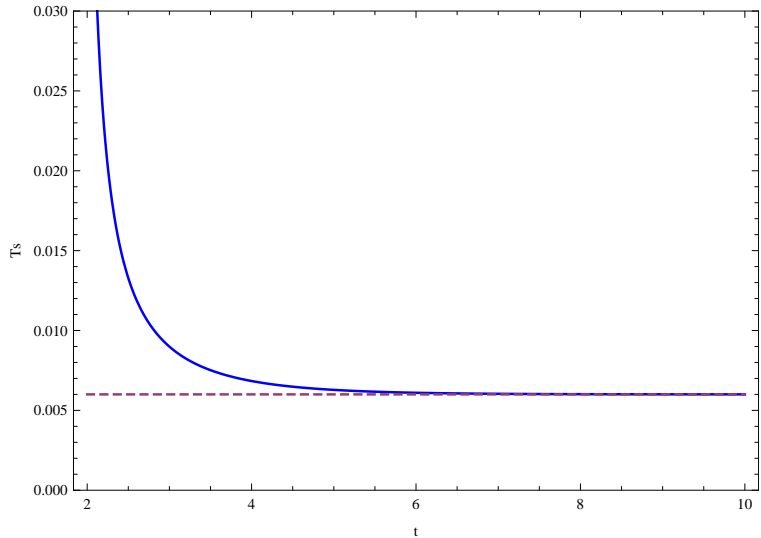}
  \caption{We plot the entropy density $s$ in terms of cosmic time $t$ where, $\Pi=0$ by dashed line and $\Pi=-3\zeta_1\rho H$ by blue curve ($T=5.47\times 10^{-5}, V_0=10^{-9}, \lambda=1, \gamma=\frac{4}{3}, \phi_i=0 , \alpha_1=6.25\times10^2,\zeta_1=3.9\times10^{-10},\alpha=8\pi$)}
 \label{fig:F2}
\end{figure}
From Eq.(\ref{60}) the scalar field and effective potential at the end of inflation where $\tilde{\epsilon}\simeq 1$, becomes
\begin{equation}\label{65}
\phi_f=\frac{1}{\alpha}\ln[\frac{\alpha_1 V_0}{(6\lambda)^{\frac{1}{2}}\alpha^2}]~~~~~~~V_f=\frac{(6\lambda)^{\frac{1}{2}}\alpha^2}{\alpha_1}
\end{equation}

so, by using the above equation and Eq.(\ref{61}) we may find time at which inflation ends
\begin{equation}\label{}
t_f=-\frac{\alpha_1}{\alpha^2}\ln[\frac{\alpha V_0}{(6\lambda)^{\frac{1}{2}}\alpha^2}]+\frac{\alpha_1}{\alpha}\phi_i
\end{equation}

Number of e-folds in this case is related to $V_i$ and $V_f$ by using Eq.(\ref{15})
\begin{equation}\label{63}
V_i=(N+1)V_f
\end{equation}

At the begining of the inflation $r$ parameter is given by
\begin{equation}\label{64}
r=r_i=\frac{\alpha_1}{3(N+1)\alpha}
\end{equation}

so, high dissipative condition($r\gg 1$), leads to $\alpha_1\gg\alpha(N+1)$.

By using  Eqs.(\ref{37}) and (\ref{45}) scalar power spectrum and tensor-scalar ratio result to be
\begin{eqnarray}\label{67}
\delta_H^2=\frac{128\sqrt{\alpha_1}}{25\sqrt[4]{6\lambda}\alpha^2}V^{2A}(\phi_0)\exp(-\frac{2B}{V(\phi_0)})\sqrt{V(\phi_0)}T_r
\end{eqnarray}

and
\begin{eqnarray}\label{68}
R=\frac{\alpha^2}{64\pi^2\sqrt{\alpha_1}\sqrt[4]{(6\lambda)^3}}V^{-2A}(\phi_0)\exp(\frac{2B}{V(\phi_0)}) \frac{V(\phi_0)^{\frac{3}{2}}}{T_r}\coth[\frac{k}{2T}]
\end{eqnarray}

respectively, where $A=\frac{9}{4}\frac{1+\frac{\zeta_1\alpha^2}{\sqrt{\alpha\lambda}\alpha_1\gamma}}{1-\frac{3}{4}\frac{\zeta_1\alpha^2}{\gamma^2\alpha_1}}$ and $B=\frac{3}{4}\frac{(\gamma-1)\alpha^2\sqrt{6\lambda}}{\gamma\alpha_1(1-\frac{3}{4}\frac{\zeta_1\alpha^2}{\gamma^2\alpha_1})}$.
In the above equation we have used the Eq.(\ref{38}) where
\begin{equation}\label{69}
\tilde{\Im}(\phi_0)=-A\ln(V(\phi_0))+\frac{B}{V(\phi_0)}
\end{equation}

We may restrict  these parameters, using WMAP7 observational data \cite{6}. Using WMAP7 data, $P_R(k_0)=\frac{25}{4}\delta_H^2\simeq 2.28\times 10^{-9}$, $R(k_0)\simeq 0.21$ and the characteristic of warm inflation $T>H$ \cite{3}, we may restrict the values of temperature to $T_r>5.47\times 10^{-5}M_4$ using Eqs.(\ref{37}), (\ref{45}), or the corresponding equations (\ref{58}), (\ref{59}), (\ref{67}), (\ref{68}) in our coupled examples, (see FIG.3). We have chosen $k_0=0.002 Mpc^{-1}$ and $T\simeq T_r$. Note that, because of the bulk viscous pressure, the radiation energy density in our model increases. Therefore the minimum value of temperature for our model ($5.47\times 10^{-5}M_4$) is bigger than the minimum value of temperature ($3.42\times 10^{-6}M_4$ ) for the model without the viscous pressure effects \cite{6-f}.
\begin{figure}[h]
\centering
  \includegraphics[width=10cm]{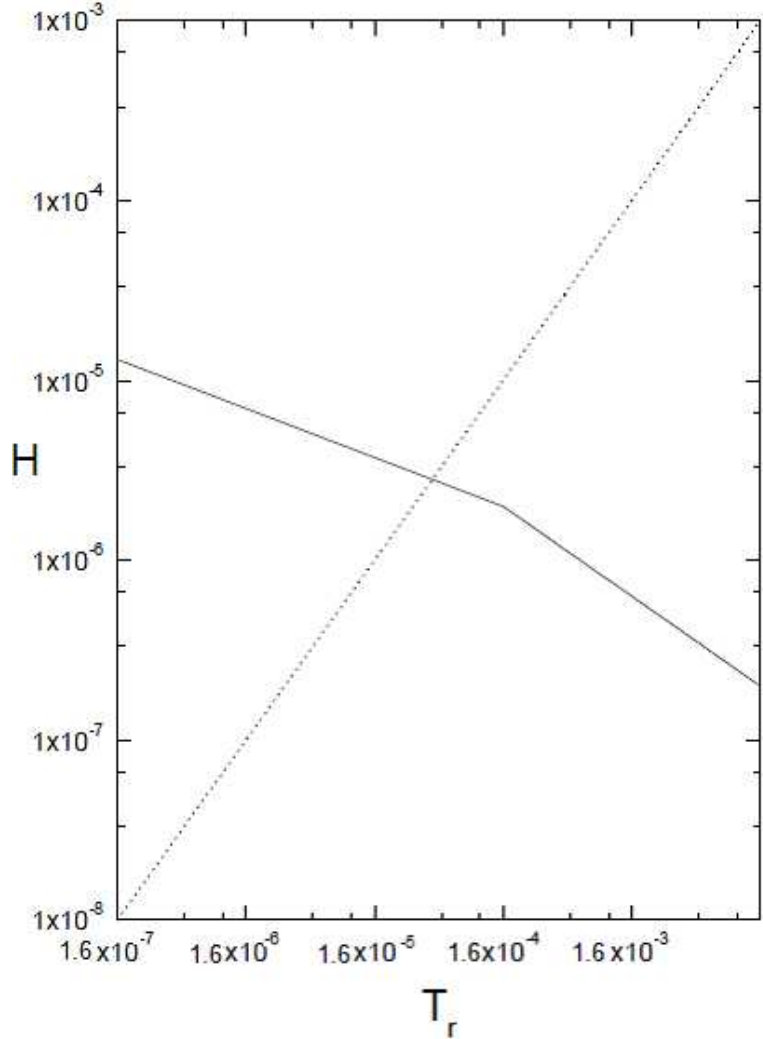}
  \caption{In this graph we plot the Hubble parameter $H$ in term of the temperature $T_r$. We can find the minimum amount of temperature $T_r=5.47\times 10^{-5}$ in order to have the necessary condition for warm inflation model ($T_r>H$). }
 \label{fig:F3}
\end{figure}
\section{Conclusion}
Warm-tachyon  inflation model with viscous pressure on the brane, using overlasting form of potential
$V(\phi)=V_0\exp(-\alpha\phi)$ which agrees with tachyon
potential properties have been studied. The main problem of the
cool inflation theory is how to attach the universe to the end of the
inflation epoch. One of the solutions for  this problem is the
study of inflation epoch in the context of warm inflation theory \cite{3}. In
this model of inflation, radiation is produced during inflation period. Production of radiation  is phenomenologically
recognized  by introducing the dissipation term $\Gamma$ in equations of motion.
Warm inflation model with viscous pressure is an extension of warm inflation model
where instead of radiation field we have radiation-matter fluid.
The study of warm inflation model with viscous pressure on the brane as a mechanism that gives an end for
tachyon inflation are motivated us to consider the warm tachyon
inflation model on the brane with viscous pressure. In this article we have considered warm-tachyon
inflationary universe  with viscous pressure on the brane. In slow-roll approximation the general relation between the energy density of radiation-matter fluid and the energy density of tachyon field is found. In longitudinal gauge and slow-roll limit the explicit expressions
for the tensor-scalar ratio $R$, scalar spectrum $P_R$, scalar spectral index
$n_s$ and its running $\alpha_s$ have been obtained. We have
developed our specific model by exponential potential for two cases: 1- Constant dissipation coefficient $\Gamma_0$ and constant bulk viscous pressure coefficient $\zeta_0$. 2- $\Gamma$ as a function of tachyon field $\phi$ and $\zeta$ as a function of imperfect fluid energy density $\rho$. In these two
cases we have found perturbation parameters and constrained these
parameters by WMAP7 observational data.
\section{Acknowledgements}
We would like to thank I. Antoniadis for bringing our attention to the last references in [13], also we would like to thank  Y. Cai  and J. Rosa for bringing our attention to the papers in Ref.[5], and their useful comment and suggestions.
Also we thank the referee for
good observations and kind guidelines.
\section{appendix A}

In this article we study our model in natural units  where $c=\hbar=1$, so we have ($[Mass]=M,[Time]=T,[Length]=L$)
\begin{eqnarray}\label{}
\nonumber
[c]=LT^{-1}=1~~~\Rightarrow ~~~~~~~T=L~~~~~~\\
\nonumber
[\hbar]=ML^{2}T^{-1}~\Rightarrow~~~~~L=T=M^{-1}
\end{eqnarray}
In Eq.(\ref{i1}) (which is obtained from Eq.(\ref{1})) we have
\begin{eqnarray}\label{}
\nonumber
[H^2]=[\frac{8\pi}{3M_4^2}\rho_T]~~\Rightarrow~~~~\frac{[a]^2}{T^2[a]^2}=\frac{[\rho_T]}{[M_4^2]}\\
\nonumber
\Rightarrow~~~[\rho_T]=[T_{\mu\nu}]=M^4~~~~~~~~~~~~~~~~~~~
\end{eqnarray}
It  is appear that $T_{\mu\nu}$ have mass dimension ~$4$.\\
 From Eqs.(\ref{2}) and (\ref{3}), we have $-\dot{\phi}^2$ plus dimensionless number ~$1$, therefore
\begin{eqnarray}\label{}
\nonumber
[\dot{\phi}]=[1]~~~\Rightarrow~~~\frac{[\phi]}{T}=[\phi]M=1
\end{eqnarray}
It is appear that tachyon field have mass dimension ~$-1$ (which is agree with the tachyonic form of potential (\ref{46}))\\
 In Eq.(\ref{6}) RHS and LHS have mass dimension ~5.
\begin{eqnarray}\label{}
\nonumber
[\Gamma\dot{\phi}^2]=[\Gamma][\dot{\phi}]^2=M^5~~~~~~~~~~~~~~~~~~~~~~~~~~~~~~~\\
\nonumber
[\dot{\rho}]=\frac{[\rho]}{T}=M^4\times M=M^5~~~~~~~~~~~~~~~~~~~~~~~\\
\nonumber
[3HP]=\frac{[\dot{a}]}{[a]}[P]=\frac{[a]}{[a]T}[P]=M\times M^4=M^5
\end{eqnarray}\\
 In Eq.(\ref{8}) RHS and LHS have mass dimension ~$1$.
\begin{eqnarray}\label{}
\nonumber
[\frac{V'}{V}]=\frac{[V]}{[V][\phi]}=M~~~~~~~~~~~~~~~~~~~\\
\nonumber
[3H\dot{\phi}]=[H][\dot{\phi}]=M~~~~~~~~~~~~~~~~~\\
\nonumber
[3Hr\dot{\phi}]=\frac{[H][\Gamma][\dot{\phi}]}{[H][V]}=\frac{M^5}{M^4}=M~~
\end{eqnarray}

\section{appendix B}
The explicit forms for the dissipation coefficient have been derived from  analysing intermediate particle production \cite{A1,A2,A3}($\phi\rightarrow \chi\rightarrow yy$, where $\chi$ is heavy intermediate field and $y$ is light field.). We follow the thermal field theory methods  which is found in Ref.\cite{A4}. Schwinger-Keldeysh approach will be used here \cite{A5,A6,A7,A8,A9,A10,A11,A12,A13}. From thermal field theory methods the ensemble of an operator $\widehat{A}$ in Heisenberg picture may be written

\begin{equation}\label{}
<\hat{A}>=\frac{\rho_{ii'}<\psi_i|\hat{A}|\psi_{i'}>}{\rho_{ii'}<\psi_i|\psi_{i'}>}=\frac{tr(\rho\hat{A})}{tr(\rho)}
\end{equation}
where $\rho$ is density matrix. Shifted version of Schwinger-Keldysh, or closed approach path formalism may be used, to evaluate ensemble average \cite{A14}. The generating functional for the scalar field $\phi$ is given by
\begin{equation}\label{}
Z[\phi_1,\phi_2,J_1,J_2]=\rho_{ii'}<\psi_i\mid T^{*}\exp(-i\int J_2(\hat{\phi}-\phi_2))T\exp(i\int J_1(\hat{\phi}-\phi_1))\mid\psi_{i'}>
\end{equation}
where $T$ ($T^{*}$) denotes time ordering with smallest time on right (left). The index $a$ for pairs $J_{a}=(J_1,J_2)$ and $\phi_a=(\phi_1,\phi_2)$ raises by a metric $c_{ab}=diag(1,-1)$ \cite{A15}. Two-point functions of this model are
\begin{equation}\label{}
G_{ab}=-\frac{i}{Z}\frac{\delta^2 Z}{\delta J^a\delta J^b}\mid_{(J_1,J_2)=0}
\end{equation}
or
\begin{equation}\label{}
G^{>}(x,x')=G^{<}(x',x)=i<\hat{\eta}(x)\hat{\eta}(x')>
\end{equation}
where $\hat{\eta}=\hat{\phi}-\phi_a$. The starting point of perturbation theory for a system close to equilibrium is free theory with thermal distribution of states. Feynman diagram expansions may be used for studying the correlation functions of interacting theory. Free theory propagators $-iG_{0ab}$ is represented by lines and vertices are found from an interacting lagrangian \cite{A5}. The study of the background fields may be identified with ensemble averages of the field operators. The equation of background field is given by
\begin{equation}\label{a1}
\frac{\delta Z}{\delta J^a}=0
\end{equation}
This equation may be solved for $J_a[\phi_b]$. From Eq.(\ref{a1}) the effective field equation is
\begin{equation}\label{a2}
F[\phi](x)=-\frac{\delta S[\phi_a]}{\delta\phi_b}\mid_{\phi_a=\phi}=0
\end{equation}
where the effective action is
\begin{equation}\label{}
S[\phi]=-i\ln Z[\phi_a,J_b[\phi_c]]
\end{equation}
We consider the tachyonic fluid  as an interacting system in thermal equilibrium. We assume the background fields are constant $\phi_1=\phi_2$ and $G_{ab}(x,x')\equiv G_{ab}(x-x')$. We define the vacuum polarization or self-energy $\Pi_{ab}$ as
\begin{equation}\label{}
G^{-1ab}(P)=(P^2+m^2)c^{ab}+\Pi^{ab}
\end{equation}
where
\begin{equation}\label{}
G_{ab}(P)=\int d^4x G_{ab}(x,x')e^{-i P.(x-x')}
\end{equation}
and $\mathbf{P}$=($\overrightarrow{P},\omega$) is 4-momentum. In FIG.4, we show diagrammatic representation for two-point function. We define the dissipative part of self-energy as
\begin{equation}\label{}
\beta=i(\Pi_{21}-\Pi_{12})
\end{equation}
where $\Pi_{ab}=-\Pi_{ab}^{*}$.

 \begin{figure}[h]
\centering
  \includegraphics[width=10cm]{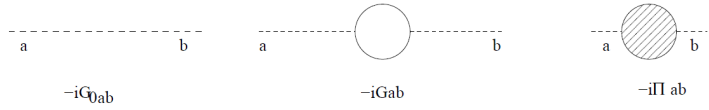}
  \caption{We plot the diagrams of vacuum energy $-i\Pi_{ab}$, two-point function $-iG_{ab}$ and free two-point function $-iG_{0ab}$. }
 \label{fig:F1}
\end{figure}

From above definitions and using the methods of Ref.\cite{A5}, we will study the effective field equation of tachyon field which is interacting with radiation field in slow-roll limit. From Eq.(\ref{a2}) the effective field equation for our model is
\begin{equation}\label{}
F(x)=-\frac{\delta S}{\delta\phi_1(x)}\mid_{\phi_a=\phi}=0
\end{equation}
where the homogeneous field $\phi$ varies slowly slowly about $\phi_t=\phi(t)$. An expansion of $F$ is given by
\begin{equation}\label{nnn1}
F(x)=\sum_{n=0}^{\infty} F_n(x)
\end{equation}
where
\begin{equation}\label{}
F_n(x)=-\frac{1}{n!}\int d^4x_1...d^4x_n\frac{\delta^{n+1}S}{\delta\phi_1(x)\delta\phi_{a_1}(x_1)...\delta\phi_{a_n}(x_n)}\mid_{\phi_a=\phi_1}\delta\phi_{a_1}(x_1)...\delta\phi_{a_n}(x_n)
\end{equation}
and $\delta\phi_a=\phi_a-\phi_t$.
The first term $F_0$ may be expressed as the derivative of $\ln(V)$. This term represents the part of field equation without time derivative terms. Term $F_1$ contains the equilibrium self-energy $\Pi_{ab}^{\eta}(k,t-t_1)$ of perturbation $\eta$ and derivative terms \cite{A5}.
\begin{equation}\label{nnn2}
F_1(x)=\ddot{\phi}+3H\dot{\phi}-\int_{-\infty}^{t}dt_1\Pi_{1a}^{\eta}(0,t-t_1)\delta\phi^a(t_1)\mid_{\phi_a=\phi}
\end{equation}
From Eqs.(\ref{nnn1}) and (\ref{nnn2}) we have
\begin{eqnarray}\label{}
F\simeq F_0+F_1=0~~~~~~~~~~~~~~~~~~~~~~~~~~~~~~~~~~~~~~~~~~~~~~~~~~\\
\nonumber
\Rightarrow~~~~~\ddot{\phi}+3HV\dot{\phi}+\frac{V'}{V}+-\int_{-\infty}^{t}dt_1\Pi_{1a}^{\eta}(0,t-t_1)\delta\phi^a(t_1)\mid_{\phi_a=\phi}=0
\end{eqnarray}
In slow-roll limit the tachyon field may be expressed as
\begin{equation}\label{}
\phi(t_1)=\phi(t)+(t_1-t)\dot{\phi}(t)+...
\end{equation}
The tachyon field equation of motion with linear dissipative term is
\begin{equation}\label{}
\ddot{\phi}+(3HV+\Gamma)\dot{\phi}+\frac{V'}{V}=0
\end{equation}
So we obtain
\begin{equation}\label{a3}
\Gamma=2\int_0^{\infty}dt'Re(\Pi^{\eta}_{21}(0,t'))t'
\end{equation}
This result may be found from linear response theory \cite{A16}. We may represent an adiabatic approximation for massive boson field $\chi$ with an interacting Lagrangian \cite{A10}
\begin{equation}\label{}
\mathfrak{L}_{I}=-\frac{1}{2}g^2\hat{\Phi}(\phi)\chi^2
\end{equation}

\begin{figure}[h]
\centering
  \includegraphics[width=10cm]{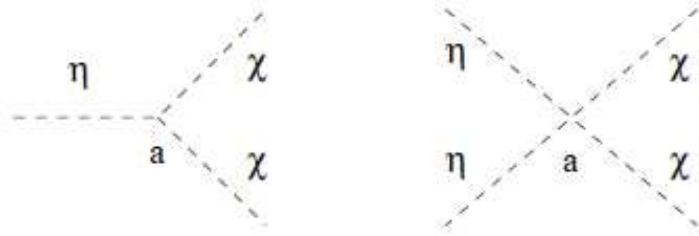}
  \caption{We plot the diagrams of vertices for interaction between inflaton $\phi$ and intermediate particle $\chi$.}
 \label{fig:F1}
\end{figure}

\begin{figure}[h]
\centering
  \includegraphics[width=10cm]{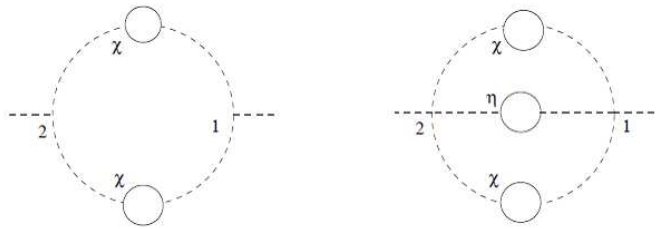}
  \caption{The $\eta$ self-energy diagrams of order $g^4$ (right) and $g^2$ (left) are plotted. }
 \label{fig:F1}
\end{figure}

In FIG.5, we show the vertices for interactions between $\eta$ and $\chi$. $\chi$ is heavy field with mass $m_{\chi}\sim\hat{\Phi}^{\frac{1}{2}}$. From left diagram in FIG.6 and Eq.(\ref{a3}), we can derive the contribution to the self energy of the field $\eta$ at order $g^2$

\begin{equation}\label{}
\Gamma=4g^4\hat{\Phi}Im\int\frac{d^3k}{(2\pi)^3}\int_0^{\infty}dt'(G_{21}^{\chi})^2t'
\end{equation}
where \cite{A5}
\begin{equation}\label{}
G_{21}^{\chi}=\int_{-\infty}^{\infty}\frac{d\omega}{2\pi}i(1+n)\rho_{\chi}e^{-i\omega t'}
\end{equation}
For Chaotic potential $\hat{\Phi}\sim V(\phi)\sim \phi^2$, we have
\begin{equation}\label{}
\Gamma\sim\phi^2
\end{equation}
which is agree with the result of Ref.\cite{A5}. In our model for tachyon potential we set $\hat{\Phi}\sim V(\phi)\sim \exp(-\alpha\phi)$. Therefore, we have
\begin{equation}\label{}
\Gamma\sim\exp(-\alpha\phi)\sim V(\phi)
\end{equation}
In section.4 we have used the $\Gamma=\alpha_1 V(\phi)$ form for dissipative coefficient.

\end{document}